# Performance of Hybrid Concatenated Trellis Codes CPFSK with Iterative Decoding over Fading Channels

Labib Francis Gergis

Misr Academy for Engineering and Technology

Mansoura, Egypt

IACSIT Senior Member, IAENG Member

*drlabeeb@yahoo.com*

*Abstract:*

*Concatenation is a method of building long codes out of shorter ones, it attempts to meet the problem of decoding complexity by breaking the required computation into manageable segments.*
*Concatenated Continuous Phase Frequency Shift Keying (CPFSK) facilitates powerful error correction. CPFSK also has the advantage of being bandwidth efficient and compatible with non-linear amplifiers.*
*Bandwidth efficient concatenated coded modulation schemes were designed for communication over Additive White Gaussian noise (AWGN), and Rayleigh fading channels. An analytical bounds on the performance of serial concatenated convolutional codes (SCCC), and parallel concatenated convolutionalcodes (PCCC), were derived as a base of comparison with the third category known as hybrid concatenated trellis codes scheme (HCTC). An upper bound to the average maximum-likelihood bit error probability of the three schemes were obtained. Design rules for the parallel, outer, and inner codes that maximize the interleaver's gain were discussed. Finally, a low complexity iterative decoding algorithm that yields a better performance is proposed.*

## I. Introduction

The channel capacity unfortunately only states what data rate is theoretically possible to achieve, but it does not say what codes to use in order to achieve an arbitrary low BER for this data rate. Therefore, there has traditionally been a gap between the theoretical limit and the achievable data rates obtained using codes of a manageable decoding complexity. However, a novel approach to error control coding revolutionized the area of coding theory. The so-called turbo codes [1], almost completely closed the gap between the theoretical limit and the data rate obtained using practical implementations. Turbo codes are based on concatenated codes separated by interleavers. The concatenated codes can be decoded using a low-complexity iterative decoding algorithm. Given certain conditions, the iterative decoding algorithm performs close to the fundamental Shannon capacity [2]. In general, concatenated coding provides longer codes yielding significant performance improvements at reasonable complexity investments. The overall decoding complexity of the iterative decoding algorithm for a concatenated code is lower than that required for a single code of the corresponding performance

Interest in code concatenation has been renewed with the introduction of turbo codes [3], otherwise known as parallel concatenated convolutional codes (PCCCs) [4,7], and the closely related serially concatenated convolutional codes (SCCCs) [5, 6], and [8]. In this paper, we introduce the parallel and serial concatenation codes as a references to compare with HCTC [9,10]. These codes perform well and yet have a low overall decoding complexity.

Previous researches had considered the spectral efficiency characteristics of several modulation schemes [11-17]. It can be shown that Continuous Phase Frequency Shift Keying (CPFSK) can achieve spectral efficiency. The system model for CPFSK in AWGN is introduced in [11], with a brief review of the symmetric information rate of CPFSK in AWGN.

Study of adaptive coded modulations for wireless channels nonlinearity due to the radio-frequency power amplifier was considered and continuous-phase modulations (CPM) are adopted in order to make the nonlinearity effects negligible [13].

The definition based on the spacing between adjacent carriers in a frequency division multiplexed CPM system, was considered with the inter-channel interference, which depends on the channel spacing in [14]. The spectral efficiency achievable by a single-user receiver in the considered multi-channel scenario was evaluated, to optimize the channel spacing with the aim of maximizing the spectral efficiency, showing that impressive improvements with respect to the spectral efficiencies.

A system involving the multisymbol noncoherent reception of CPFSK was developed, optimized, and analyzed in [15], the achievable performance over AWGN and Rayleigh block fading channels was determined by computing the average mutual information, which is the capacity of a channel using the given modulation format and receiver architecture under the constraint of uniformly distributed input symbols.

Digital modulation schemes with channel coding was closely associated to make up the heart of the physical layer of telecommunication systems in [16]. An error performance prediction model for a modulated concatenated turbo coded link was proposed in [17]. This model predicted performance addressing the fading phenomena for wireless radio channels.

In addition to its spectral characteristics, CPFSK possesses two other appealing characteristics. First, CPFSK maintains a constant amplitude signal, which is appropriate for nonlinear channels, as it will experience fewer adverse effects than a non-constant envelope signal. Hence, a non-linear high power amplifier in the signal path is acceptable. Second, the information in a CPFSK signal can be retrieved via non-coherent demodulation, which is appropriate for multipath fading channels.

This paper is organized as follows. Section II describes Continuous Phase Frequency Shift Keying (CPFSK). Section III describes in details the system model and encoder structure of hybrid concatenated code (HCTC), derives an analytical upper bounds to the bit-error probability of HCTC using the concept of uniform interleavers that decouples the output of the outer encoder from the input of the inner encoder, and the input of the parallel code from the input of the outer code. Factors that affect the performance of HCTC are described in section IV. Finally conclusion results for some examples described in section III, have been considered in section V.

## II. CPFSK System Model

Continuous Phase Frequency Shift Keying [11], is quite often used for modulation in modern short-range wireless systems, e.g. according to the Bluetooth standard. The reason for this is its inherent noise immunity and the possibility to use high efficiency non-linear power amplifiers on the transmitter side. Due to the advancement of digital signal processing receiver architectures with very low intermediate frequencies (IF) became more and more popular in the last years.

In general CPFSK, the signals of two different frequencies of $f_0$, and $f_1$ to transmit a message $m=0$ or $m=1$, over a time of $T_b$ seconds is transmitted as

$$S_o(t) = \sqrt{2E_b / T_b} \cos \{ 2\pi f_0 t + \theta(0) \} \qquad 0 \leq t \leq T_b$$
$$S_1(t) = \sqrt{2E_b / T_b} \cos \{ 2\pi f_1 t + \theta(0) \} \qquad 0 \leq t \leq T_b \qquad (1)$$

where $E_b$ is the transmitted signal energy per bit,
$T_b$ is the bit duration, and
$\theta(0)$, the value of the phase at $t=0$.

Equation (1) can be rewritten as

$$S_o(t) = \sqrt{2E_b / T_b} \cos \{ 2\pi f_c t + \theta(t) \} \qquad (2)$$

where $\theta(t)$ is the phase of $S(t)$ given by

$$\theta(t) = \theta(0) \pm \pi h / T_b \qquad 0 \leq t \leq T_b \qquad (3)$$

Minimum shift keying (MSK) is a special type of continuous phase-frequency shift keying (CPFSK) with a modulation index $h=0.5$, corresponds to the minimum frequency spacing, $f_0 - f_1 = 1/(2T_b)$, that allows two FSK signals to be coherently orthogonal, and the name minimum shift keying implies the minimum frequency separation (i.e. bandwidth) that allows orthogonal detection.

## III. Performance of Hybrid Concatenated Trellis Codes

Consider an $(n, k)$ block code $C$ with code rate $R_c = k / n$ and minimum distance $h_{min}$. An upper bound on the conditional bit-error probability of the block code $C$ over fading channels, assuming coherent detection, maximum likelihood decoding, and perfect Channel State Information (CSI) can be obtained in the form [10]

$$P_b(e/\rho) \leq \sum_{h=h_{min}}^{n} \sum_{w=1}^{k} (w/k) \, A^c_{w,h} \, Q \left( \sqrt{2 R_c(E_b/N_o) \sum_{i=1}^{h} \rho^2_i} \right) \qquad (4)$$

where $E_b / N_o$ is the bit energy-to-noise density ratio per bit, and $A^c_{w,h}$ for block code $C$ represents the number of codewords of the block code with output weight $h$ associated with an input sequence of weight $w$. The $A^c_{w,h}$ is the input-output weight coefficient (IOWC). The function $Q \sqrt{2R_c(E_b/N_o) \sum \rho^2_i}$ represents the pairwise error probability which is a monotonic decreasing function of the signal to noise ratio and the output weight

*h*. The fading samples *ρ* are independent identically distributed (*i. i. d.*) random variables with *Rayleigh* density of the form

$$f(\rho) = 2\rho e^{-\rho^2} \quad (5)$$

The *Q* function can be represented as

$$Q(x) \leq (1/2) e^{-x/2}$$

The structure of (HCTC) is shown in Fig. 1. It is composed of three concatenated codes: the *parallel* code $C_p$ with rate $R^p_c = k_p / n_p$ and equivalent block code representation $(N_1/R^p_c, N_1)$, the outer code $C_o$ with rate $R^o_c = k_o / p_o$ and equivalent block code representation $(N_1/R^o_c, N_1)$, and the inner code $C_i$ with rate $R^i_c = p_i / n_i$ and equivalent block code representation $(N_2/R^i_c, N_2)$. With two interleavers $\Pi_1$ and $\Pi_2$ have $N_1$ and $N_2$ bits long, generating a hybrid turbo trellis code $C_H$ with overall rate $R_H$. For Simplicity, we assume $k_p = k_o$ and $p_o = p_i = p$.
This gives a HCTC scheme with overall rate $R_H = k_o / (n_p + n_i)$.

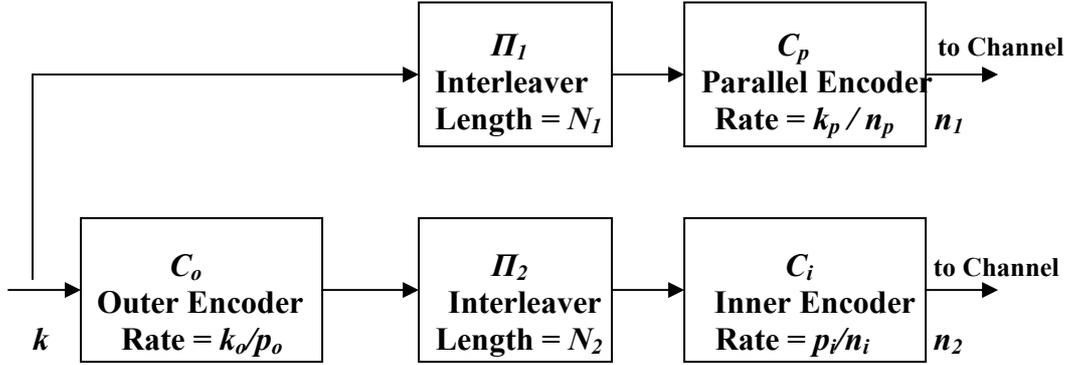

Fig. 1. A Hybrid Concatenated TrellisCode (HCTC)

Since the (HCTC) has two outputs, the upper bound on the bit-error probability in equation (4) can be modified to

$$P_b(e/\rho) \leq \sum_{h_1=h^p_m}^{n_1} \sum_{h_2=h^i_m}^{n_2} \sum_{w=w_m}^{k} (w/k) A^{cH}_{w,h_1,h_2} Q\left[\sqrt{2R_H (h_1+h_2)(E_b/N_o)}\right] \quad (6)$$

where $A^{cH}_{w,h1,h2}$ for the (HCTC) $C_H$ represents the number of codewords of the equivalent block code with output weight $h_1$ for the parallel code and output weight $h_2$ for the inner code associated with an input sequence of weight w, $A^{cH}_{w,h1,h2}$ is the IOWC for the HCTC, $w_m$ is the minimum weight of an input sequence generating the error events of the parallel code and the outer code, $h^p_{min}$ is the minimum distance of the parallel encoder $C_p$, and $h^i_{min}$ is the minimum distance of the inner encoder $C_i$. Using the properties of uniform interleavers, the first interleaver transforms the input data of weight *w* at the input of the outer encoder into all its distinct $\binom{N_1}{w}$ permutations at the input of the parallel encoder.

Similarly, the second interlaver transforms a codeword of weight *l* at the output of the outer encoder into all its distinct $\binom{N_2}{l}$ permutation at the input of the inner encoder.

Thus the $A^{CH}_{w,h1,h2}$ (IOWC) for the HCTC, can be obtained as [9],[10]

$$A^{C_H}_{w,h1,h2} = \sum_{l=0}^{N_2} \frac{A^{C_p}_{w,h1} \times A^{C_o}_{w,l} \times A^{C_i}_{l,h2}}{\binom{N_1}{w}\binom{N_2}{l}} \qquad (7)$$

where $A^{C_o}_{w,L}$ is the number of codewords of the outer code of weight *l* given by the input sequences of weight *w*.

Let $A^{C}_{w,h,j}$ be the input-output weight coefficients given that the convolutional code generates *j* error events with total input weight *w* and output weight *h*. The $A_{w,h,j}$ actually represents the numbers of sequences of weight *h*, input weight *w*, and the number of concatenated error events *j* without any gaps between them, starting at the beginning of the block. For *N* much larger than the memory of the convolutional code, the coefficient $A^{C}_{w,h}$ of the equivalent block code can be approximated by

$$A^{C}_{w,h} \approx \sum_{j=1}^{n_M} \binom{N/p}{j} A^{C}_{w,h,j} \qquad (8)$$

where $n_M$, is the largest number of error events concatenated in a codeword of weight *h* and generated by weight *w* input sequence, is a function of *h* and *w* that depends on the encoder.

Using expression (8) for the case of (HCCC) with *j* replaced by $n^i$ for the inner code, *j* replaced by $n^o$ for the outer code, and, similarly, *j* replaced by $n^p$ for the parallel code, and noting that $N_2/p = N_1/k \cong N$, we obtain for the outer code and similar expressions for the inner and parallel codes.

$$A^{C_o}_{w,L} \approx \sum_{n^o=1}^{n^o_M} \binom{N}{n^o} A^{C}_{w,l,no} \qquad (9)$$

Then substituting them into equation (6), we obtain the bit-error probability bound of the (HCTC) as

$$P_b(e/\rho) \leq \sum_{h_1=h^p_m}^{N_1/R^p_c} \sum_{h_2=h^i_m}^{N_2/R^i_c} \sum_{w=w_m}^{N_1} \sum_{l=0}^{N_2} \sum_{n^p=1}^{n^p_m} \sum_{n^o=1}^{n^o_m} \sum_{n^i=1}^{n^i_m} A^{p}_{w,h1,n^p} A^{o}_{w,l,n^o} A^{i}_{l,h2,n^i}$$

$$\cdot \frac{\binom{N}{n^p}\binom{N}{n^o}\binom{N}{n^i}}{\binom{N_1}{w}\binom{N_2}{l}}$$

$$\cdot (w/N_1) \; Q\left[\sqrt{2R_H(h_1+h_2)(E_b/N_o)}\right] \qquad (10)$$

The binomial coefficient has the asymptotic approximation

$$\binom{N}{n} \approx \frac{N^n}{n!}$$

Substituting of this approximation in equation (10) gives the bit error probability bound in the form

$$P_b(e/\rho) \leq \sum_{h_1=h^p_m}^{N_1/R^p_c} \sum_{h_2=h^i_m}^{N_2/R^i_c} \sum_{w=w_m}^{N_1} \sum_{l=d^o_f}^{N_2} \sum_{n^p=1}^{n^p_m} \sum_{n^o=1}^{n^o_m} \sum_{n^i=1}^{n^i_m} N^{np+no+ni-w-l-1}$$

$$B_{h1,h2,w,L,np,no,ni} \cdot Q\left[\sqrt{2R_H(h_1+h_2)(E_b/N_o)}\right] \quad (11)$$

where

$$B_{h1,h2,w,l,np,no,ni} = \frac{w!\; l!}{P^L k^w_o\; n^p!\; n^o!\; n^i!} \frac{w}{k_o} A^p_{w,h1,n^p}\; A^o_{w,l,n^o}\; A^i_{l,h2,n^i} \quad (12)$$

For large $N$, and for given $h_1$ and $h_2$, the dominant coefficient corresponding to $h_1 + h_2$ is the one for which the exponent of $N$ is maximum. Define this exponent as

$$\alpha(h_1, h_2) \cong \max\{n^p + n^o + n^i - w - l - 1\}$$

For large values of $E_b/N_o$, the performance of the HCTC is dominated by the first terms of the summations in $h_1$ and $h_2$, corresponding to the minimum values $h_1 = h^p_m$ and $h_2 = h^i_m$. Noting that $n^p_M$, $n^o_M$, and $n^i_M$ are the maximum number of concatenated error events in codewords of the parallel, outer, and inner code of weights $h^p_m$, $l$, and $h^i_m$, respectively. Using the method used in [9], we get

$$\alpha(h^p_m, h^i_m) \leq -d^o_f \quad (13)$$

where $d^o_f$ is the minimum hamming distance of the outer code. Substitution of the exponent $\alpha(h^p_m, h^i_m)$ into expression (11) truncated to the first term of the summation in $h_1$ and $h_2$ yields

$$P_b(e) \leq B_m\; N^{-d^o_f}\; Q\left[\sqrt{2R_H(h^p_m + h^i_m)(E_b/N_o)}\right] \quad (14)$$

where the constant $B_m$ is independent of $N$ and can be computed from (12).

Fig. 2. illustrates a comparison between the performance analysis of the three various types of the concatenated codes through the same length of uniform interleavers $N = 100$, overall rate = 1/2 for SCCC and PCCC. The overall rate for HCTC = 1/4 ( formed by $R^p_c = 1/2$, $R^o_c = 1/2$, and $R^i_c = 2/3$ ), over AWGN and Rayleigh fading channels, with MSK modulation scheme.

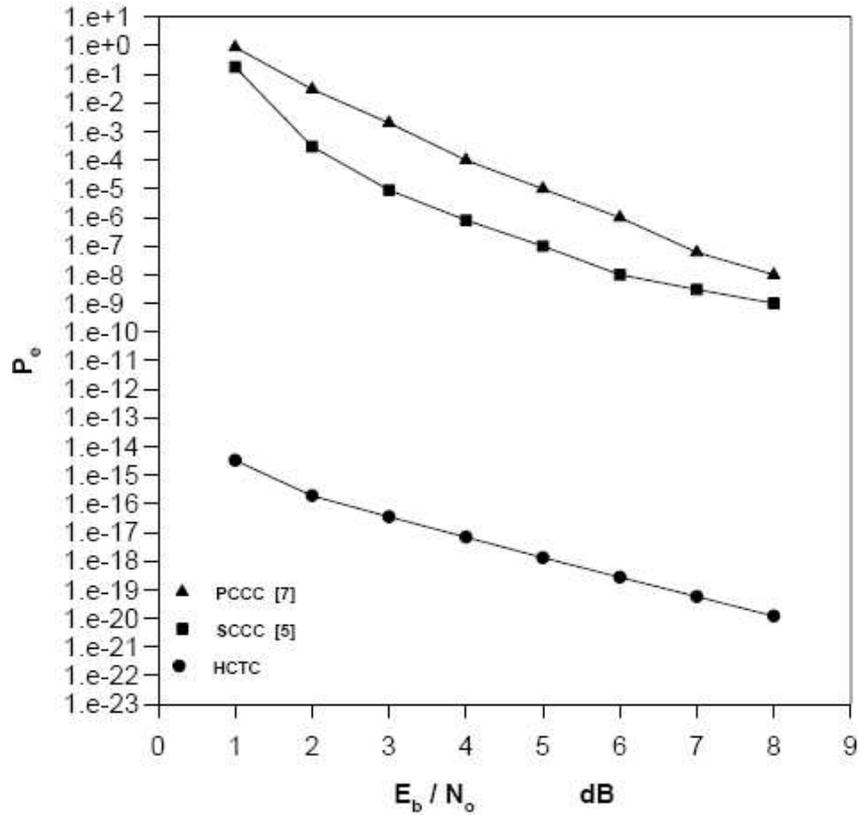

**Fig. 2. Comparison of Performance analysis for PCCC, SCCC, and HCTC over AWGN and Rayleigh Fading**

## IV.  HCTC : Performance Factors

It is shown from equation (14) that there are many factors that affect the performance of HCTC. The most influential parameter is the interleaver size $N$. As the frame of interleaver size increases, performance improves. It is shown in Fig. 3, the performance of HCTC with CPM scheme . The HCTC has an overall rate $R_H = 1/4$ formed by a parallel trellis encoder with rate *1/2*, an outer trellis code with rate 1/2, and an inner trellis code with rate *2/3*, joined by two uniform interleavers of length $N_1 = N$ and $N_2 = 2N$, where $N$ = 50, 100, 200, 300, and 500. Declaring that as the interleaver gets smaller, performance degrades.

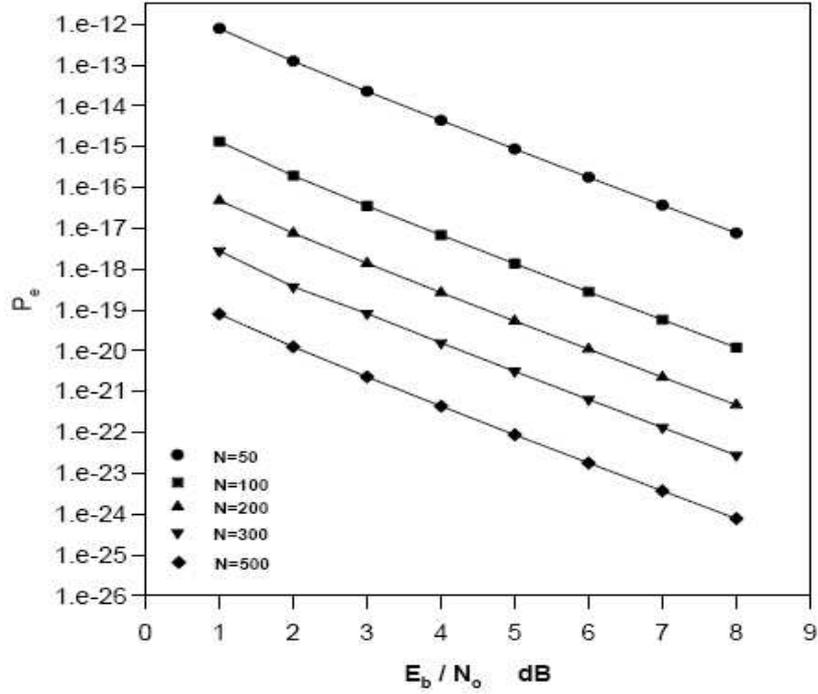

**Fig. 3. Analytical Bounds for HCTC with varying Interleaver Size $N$ over AWGN and Rayleigh Fading**

It is also clear from equation (14) that, the minimum hamming distance of the outer code ($d^o_f$) is an another main parameter in affecting the performance of HCTC. For any signal-to-noise ratios, the performance of a code is approximated by its hamming distance [1], and [2]. The design of concatenated trellis codes focuses on maximizing hamming distance, by increasing the code constraint length, that is defined to be $v$, the number of memory elements. Considering our example, the outer trellis code with rate 1/2, with constraint lengths 3, 4, 5, 6, 7 and 15 have $d^o_f$ of 6, 7, 8, 10, 12, and 18. Fig. 4. states that the bit error performance $Pe$ could be improved by increasing number of memory elements for the outer trellis encoder.

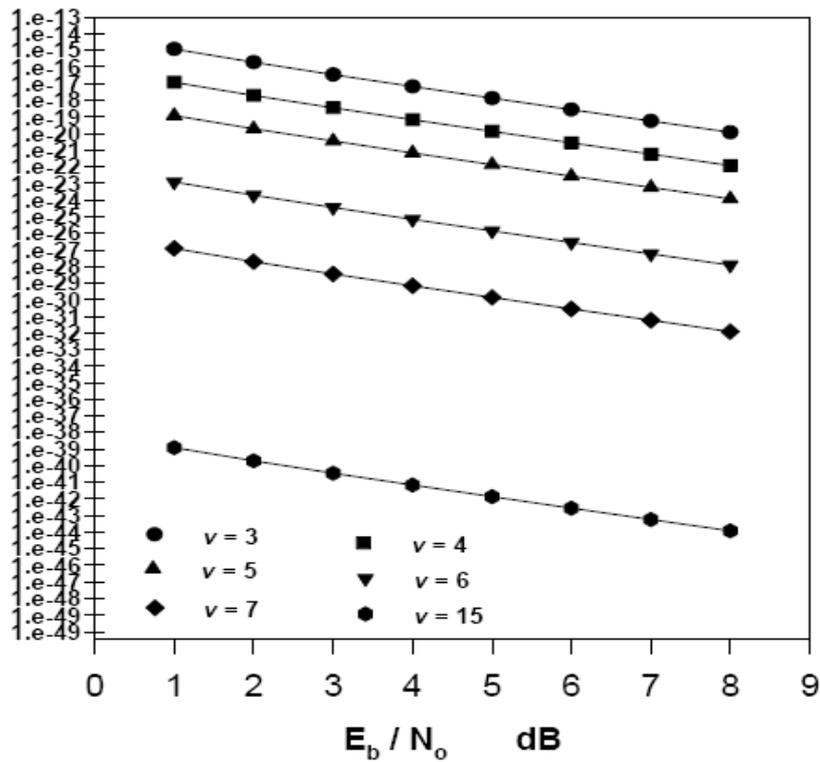

**Fig. 4. Performance of rate 1/4 HCTC with varying code constraint length over AWGN and Rayleigh Fading**

The choice of decoding algorithm and number of decoder iterations also influences performance. A functional diagram of the iterative decoding algorithm for HCTC [17], is presented in Fig. 5

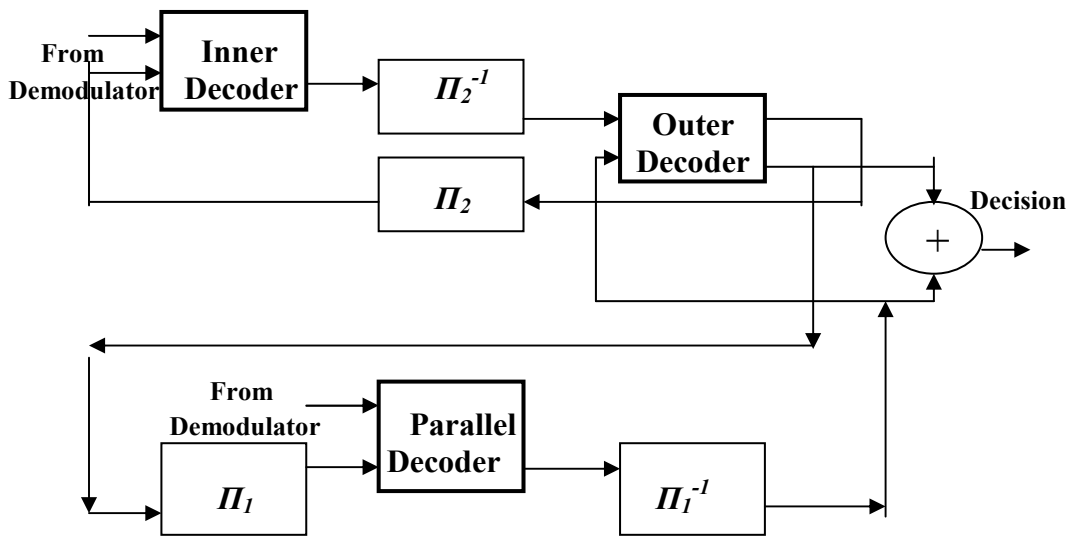

**Fig. 5. The Iterative Decoding Process for HCTC**

The performance of HCTC with MSK modulation scheme considered are shown in Fig. 6. The HCTC has an overall rate $R_H = 1/4$, the interleaver length $N$ of this code = 200 bits. The performance after various numbers of decoder is shown. It is clear that performance improves as the number of decoder iterations increases.

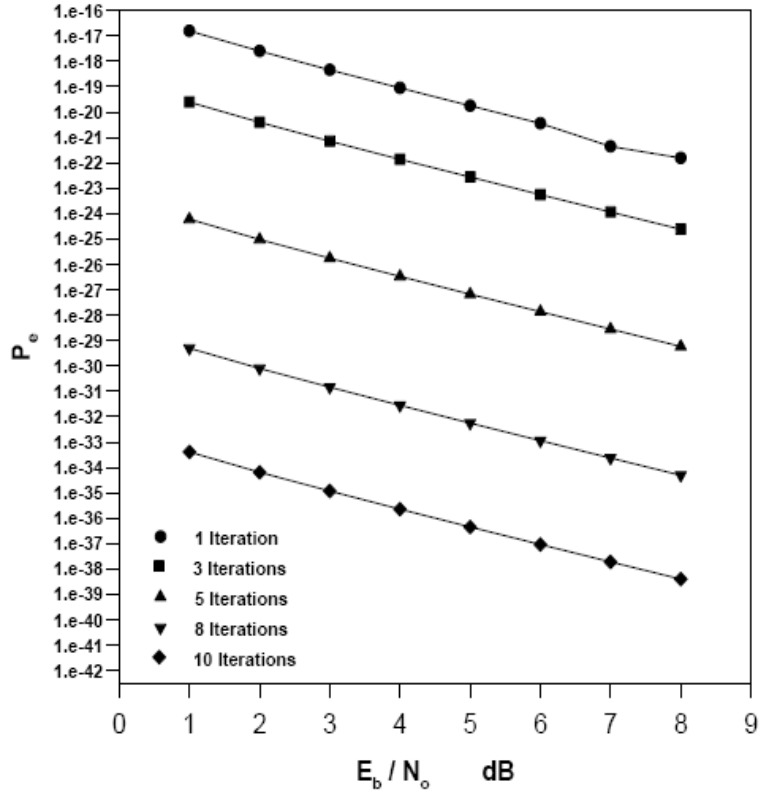

Fig. 6. Performance of rate 1/4 HCTC with various number of decoder iterations over AWGN and Rayleigh Fading

## V. Conclusions

In this paper, a number of powerful recent classes of serial and parallel concatenated trellis codes have been analyzed and compared with a proposed third choice called hybrid concatenated trellis code HCTC

A hybrid concatenated code with two interleavers is the parallel concatenation of an encoder, which accepts the permuted version of the information sequence as its input, with a serially concatenated code, which accepts the unpermuted information sequence.

These comparisons shows the superiority of HCTC over the classical SCCC and PCCC schemes. It is also demonstrated the significant in the performance and the decrease of the bit error rate and probability of errors to HCTC within increasing: the interleaver size $N$, the code constraint length, and the number of decoder iterations.

The main advantage of HCTC over SCCC, and PCCC could be seen from Fig. 2. Based on analysis was derived in section III, the interleaver size $N$. As the frame of interleaver size increases, performance improves, as it was illustrated in Fig. 3.

The bit-error probability versus the number of iterations was depicted in Fig. 6. It is clear that performance improves as the number of decoder iterations increases.

# References

[1] N. S. Muhammad, " Coding and Modulation for Special Efficient Transmission ", PhD Dissertation From the faculty of computer science, electrical engineering and information technology, Stuttgart University, 2010.

[2] M. Lahmer, M. Belkasmi, " A New Iterative Threshold Decoding Algorithm for one Step Majority Logic Decodable Block Codes", International Journal of Computer Applications, VOL. 7, No. 7, October 2010.

[3] J. Rumanek, and J. Sebesta, " New Channel Coding Methods for Satellite Communication ", RADIOENGINEERING, VOL. 19, No. 1, April 2010.

[4] B. Cristea, " Viterbi Algorithm for Iterative Decoding of Parallel Concatenated Convolutional Codes", 18 European Signal Processing Conference (EUSIPCO 2010) , Aalborg, Denemark, August 2010.

[5] R. Maunder, and L. Hanzo, " Block-Based Precoding for Serially Concatenated Codes", IEEE TRANSACTIONS ON COMMUNICATIONS LETTERS, VOL. 1, NO. 1, JANUARY 2009.

[6] R. Maunder, and L. Hanzo, " Iterative Decoding Convergence and Termination of Serially Concatenated Codes", IEEE TRANSACTIONS ON VEHICULAR TECHNOLOGY, VOL. 59, NO. 1, JANUARY 2010.

[7] F. Ayoub, M. Lahmer, M. Belkasmi, and E. Bouyakhf, " Impact of the decoder connection schemes on iterative decoding of GPCB codes", International Journal of Information and Communication Engineering, VOL. 6, No. 3, 2010.

[8] A. Farchane, and M. Belkasmi, " Generalized Serially Concatenated codes: Construction and Iterative Decoding ", International Journal of Mathematical and Computer Sciences, VOL. 6, No. 2, 2010.

[9] A. Bhise, and P. Vyavahare, " Improved Low Comlexity Hybrid Turbo Codes and Their Performance Analysis ", IEEE TRANSACTIONS ON COMMUNICATIONS, VOL. 58, NO. 6, JUNE 2010.

[10] C. Koller, A. Amai, J. Kliewer, F. Vatta, and D. Costello," Hybrid Concatenated Codes with Asymptotically Good Distance Growth", Proc. IEEE International Symposium on Turbo Codes& Related Topics, pp. 19-24, September 2008.

[11] S. Cheng, M. Valenti, and D. Torrieri, " Coherent Continuous-Phase Frequency-Shift Keying: Parameter Optimization and Code Design ", IEEE TRANSACTIONS ON WIRELESS COMMUNICATIONS, VOL. 8, NO. 4, APRIL 2009.

[12] A. Kubankova, and D. Kubanek, " Algorithms of Digital Modulation Classification and Their Verification ", WSEAS TRANSACTIONS on COMMUNICATIONS, Issue 9, Volume 9, September 2010.

[13] A. Perotti, P. Remlein, and S. Benedetto," Adaptive Coded Continuous-Phase Modulations for Frequency-Division Multiuser Systems",  ADVANCED IN ELECTRONICS AND TELECOMMUNICATIONS, VOL. 1, NO. 1, APRIL 2010.

[14] A. Barbieri, D. Fertonani, G. Colavolpe, " Spectrally-Efficient Continuous Phase

Modulations", IEEE TRANSACTIONS ON WIRELESS COMMUNICATIONS, VOL. 8, NO. 3, MARCH 2009.

[15] M. Valenti, S. Cheng, and D. Torriere, " Iterative Multisymbol Noncoherent Reception of Coded CPFSK ", IEEE TRANSACTIONS ON COMMUNICATIONS, VOL. 58, NO. 7, JULY 2010.

[16] C. Berrou, " Codes and Turbo Codes", 1 st Edition, Collection IRIS, Springer, 2010.

[17] M. Anis, " Performance Prediction of a Turbo-Coded Link in Fading Channels", Master's Thesis, AALTO UNIVERSITY SCHOOL OF SCIENCE AND TECHNOLOGY, March, 2010.